\documentclass[12pt]{article}

\usepackage{amsmath,amsfonts,amssymb,latexsym}

\newtheorem{theorem}{Theorem}[section]
\newtheorem{corollary}{Corollary}[section]
\newenvironment{proof}[1][Proof]{\textsc{#1.} }{\ \rule{0.5em}{0.5em}}
\numberwithin{equation}{section}
\def\be{\begin{equation}}
\def\ee{\end{equation}}
\def\bq{\begin{eqnarray}}
\def\eq{\end{eqnarray}}
\def\beq{\begin{eqnarray*}}
\def\eeq{\end{eqnarray*}}
\def\f{\phi}
\def\a{\alpha}
\def\b{\beta}
\def\g{\gamma}
\def\d{\delta}
\def\l{\lambda}
\def\m{\mu}

\def\na{\nabla}
\def\pa{\partial}

\begin{document}
\begin{titlepage}
\begin{flushright}
%add
\end{flushright}

\vspace{0.7cm}

\begin{center}
{\huge Slice Energy in Higher-Order Gravity Theories and Conformal Transformations}

\vspace{1cm}

{\large Spiros Cotsakis}\\

\vspace{0.5cm}

{\normalsize {\em Research Group of Mathematical Physics and Cosmology}}\\
{\normalsize {\em Department of Information and Communication Systems Engineering}}\\
{\normalsize {\em University of the Aegean}}\\
{\normalsize {\em Karlovassi 83 200, Samos, Greece}}\\
{\normalsize {\em E-mail:} \texttt{skot@aegean.gr}}
\end{center}

\vspace{0.7cm}

\begin{abstract}
\noindent We study the generic transport of slice energy between the
scalar field generated by the conformal transformation of
higher-order gravity theories and  the matter component. We give
precise relations for this exchange in the cases of dust and perfect
fluids. We show that, unless we are in a stationary spacetime where
slice energy is always conserved, in non-stationary situations
contributions to the total slice energy depend on whether or not
test matter follows geodesics in both frame representations of the
dynamics, that is on whether or not the two conformally related
frames are physically indistinguishable.
\end{abstract}

\vspace{1cm}
\begin{center}
{\line(5,0){280}}
\end{center}

\end{titlepage}

\section{Introduction}
With recent advances in observational cosmology
\cite{obs} $f(R)$ theories (and the closely related
family of scalar-tensor ones) have in the last few years regained
much  attention both in cosmology \cite{hog-cosm} and in
other contexts  \cite{hog}. Perhaps the most
economical and convenient way to study such modified gravity
theories  is through their well-known conformal relation to
general relativity, a technique  developed in the eighties
by different groups \cite{con}.

In this conformal method the $f(R)$-vacuum equations on a spacetime
$(\mathcal{V},g)$ (the so-called Jordan frame) are transformed via a
conformal transformation of the type $\tilde{g}=e^\phi g$, where
$\f$ is a function of $\pa f/\pa R$, to become on the conformally
related spacetime $(\mathcal{V},\tilde{g})$ (the so-called Einstein
frame) Einstein equations for the metric $\tilde{g}$  with the
scalar field $\f$ having a self-interacting potential that depends
on the function $f(R)$ and its first derivatives. This technique
provides a refreshing way to view the $f(R)$-vacuum theory as a
unified theory of gravitation (described in the Einstein frame by
the metric $\tilde{g}$) and the scalar field $\f$, that is as a
theory uniting general relativity and the (lagrangian) theory of the
scalar field. In the Jordan frame only one single  geometric object
appears, the `metric' $g$, and the conformal transformation then
serves as a tool to `fragment' $g$ into its two pieces in the
Einstein frame, namely the gravitational field $\tilde{g}$ and the
scalar field $\f$.

In a higher-order gravity theory with matter we have the following
different pieces of information involved in the conformal
transformation:
\begin{itemize}
\item \emph{gravity}, the field $g$ or the conformally related  field $\tilde g$
\item the scalar field $\f$.
\item  the various  matter fields $\psi$ which couple
non-minimally to  $\tilde g$ and to $\f$, or their  conformal
transform  $\tilde\psi$ which couples minimally to  $\tilde g$ but
is not coupled to $\f$.
\end{itemize}
The conformal transformation then relates the different pieces of
matter and spacetime geometry and describes the interaction between
the components listed above in the context of $f(R)$ theories.

We describe in this paper  how interactions of this sort lead to an
exchange of slice energy between the various fields and spacetime
geometry. For more  varied interactions and energy transfer models
of interest to cosmological situations see \cite{ba1, ba2} and
references therein.

The plan of this paper is as follows. The next Section is
preliminary and includes two simple applications (Corollaries 2.1,
2.2) of the basic properties (Theorems 2.1, 2.2) of the slice
energy. These basic properties and also their proofs are included
here for easy reference and also to establish notation. Their
applications deal with the simpler case of $f(R)$ theories in
vacuum. Section 3 is the heart of this paper. There, we find the
general (slice) energy transport equation and study in detail its
application to the case of a general $f(R)$ theory coupled to
matter, where by matter we mean a general  perfect fluid-scalar
field system. There are many properties analysed here but a
particularly important one deals with  conditions under which the
choice of `physical' metric influences the net contribution to slice
energy of the system. We conclude with a  discussion of
these results in Section 4.

\section{Energy on a slice}
Consider a time-oriented spacetime $(\mathcal{V},g)$ with
$\mathcal{V}=\mathcal{M} \times \mathbb{R},$ where $\mathcal{M}$
is a smooth manifold of dimension $n$, $g$ a spacetime metric and
the spatial slices $\mathcal{M}_{t}\,(=\mathcal{M}\times \{t\})$
are spacelike submanifolds endowed with the time-dependent spatial
metric $g_{t}$. (In the following, Greek indices run from $0$ to $n$,
while Latin indices run from $1$ to $n$. We also assume that the metric
signature is $(+-\cdots -)$.) On $(\mathcal{V},g)$ we consider a family of
matterfields denoted collectively as $\psi $, assume that the field $\psi$ arises
from a lagrangian density which we denote by $L$ and denote the stress tensor of
the field $\psi$  by $T(\psi )$.

For $X$ any causal vectorfield of $\mathcal{V}$ we define the
\emph{energy-momentum vector $P$ of a stress tensor $T$ relative to
$X$} to be \be\label{p} P^{\b}=X_{\a}T^{\a\b}. \ee The \emph{energy
on the slice $\mathcal{M}_{t}$ with respect to $X$}  is defined by
the integral (when it exists) \be
E_{t}=\int_{\mathcal{M}_{t}}P^{\a}n_{\a}d\m_{t}, \ee where $n$ is
the unit normal to $\mathcal{M}_{t}$ and $d\m_{t}$ is the volume
element with respect to the spatial metric $g_{t}$. We call
$P^{\a}n_{\a}$ the \emph{energy density}. Assume that $X$ and $T$
are smooth. Then we have
$$
\na_{\a}P^\a
=\na_{\a}(X_{\b}T^{\a\b})=\na_{\a}X_{\b}T^{\a\b}+X_{\b}\na_{\a}T^{\a\b}
$$
or, equivalently,
\be
\na_{\a}P^\a
=\frac{1}{2}T^{\a\b}(\na_{\a}X_{\b}+\na_{\b}X_{\a})+X_{\b}\na_{\a}T^{\a\b}.
\ee
Thus, if $\mathcal{K}\subset\mathcal{V}$ is a compact domain with
smooth boundary $\pa\mathcal{K}$, it follows from Stokes' theorem
that
\be
\int_{\mathcal{K}}\na_{\a}P^\a d\mu
=\int_{\pa\mathcal{K}}P^{\a}n_{\a}d\sigma,
\ee
where $d\mu$ is the volume element of $\mathcal{V}$ and $d\sigma$ that of $\pa\mathcal{K}$,
 and so we find
\be
\int_{\pa\mathcal{K}}P^{\a}n_{\a}d\sigma=
\frac{1}{2}\int_{\mathcal{K}}T^{\a\b}(\na_{\a}X_{\b}+\na_{\b}X_{\a})d\mu+
\int_{\mathcal{K}}X_{\b}\na_{\a}T^{\a\b}d\mu. \ee Hence, when
$\mathcal{M}$ is compact or the field falls off appropriately at
infinity, on the spacetime slab
$\mathcal{D}=\Sigma\times[t_{0},t_{1}]$,
$\Sigma\subset\mathcal{M}$, and with $T$ having support on
$\mathcal{D}$ we have the following relation for the energies on
the two end-slices \bq
\int_{\mathcal{M}_{t_{1}}}P^{\a}n_{\a}d\mu_{t_{1}}-
\int_{\mathcal{M}_{t_{0}}}P^{\a}n_{\a}d\mu_{t_{0}}
&=&\frac{1}{2}\int_{t_{0}}^{t_{1}}\int_{\mathcal{M}_{t}}
T^{\a\b}(\na_{\a}X_{\b}+\na_{\b}X_{\a})d\mu\nonumber\\
&+&\int_{t_{0}}^{t_{1}}\int_{\mathcal{M}_{t}}
X_{\b}\na_{\a}T^{\a\b}d\mu \eq or \be\label{trans-eq}
E_{t_{1}}-E_{t_{0}}=\frac{1}{2}\int_{t_{0}}^{t_{1}}\int_{\mathcal{M}_{t}}
T^{\a\b}(\na_{\a}X_{\b}+\na_{\b}X_{\a})d\mu
+\int_{t_{0}}^{t_{1}}\int_{\mathcal{M}_{t}}
X_{\b}\na_{\a}T^{\a\b}d\mu. \ee We therefore see that when $X$ is a
Killing vectorfield the first term on the right-hand-side  of Eq.
(\ref{trans-eq}) is zero and so we have \be\label{en1}
E_{t_{1}}-E_{t_{0}}=\int_{t_{0}}^{t_{1}}\int_{\mathcal{M}_{t}}
X_{\b}\na_{\a}T^{\a\b}d\mu. \ee Thus we have shown the following
result (cf. \cite{ch-mo00}, p. 87-88).
\begin{theorem}\label{1}
When $X$ is a Killing vectorfield and the field is
conserved, i.e., $\na_{\a}T^{\a\b}=0$, we have
\be
E_{t_{1}}=E_{t_{0}}.
\ee
\end{theorem}
This means that, when the energy-momentum tensor of a field
is conserved, the same is true for its slice energy relative to a Killing vectorfield
as a function of time.

In the next Section we pay particular attention to the case for
which  the field is a matter field $\psi$ interacting with a scalar
field $\phi$ with potential $V(\phi)$. We take the scalar field
lagrangian density to be \be
L=-\frac{1}{2}g^{\a\b}\pa_{\a}\f\pa_{\b}\f +V(\phi). \ee Then the
energy-momentum tensor of $\phi$ is \be\label{t} T^{\a\b}(\f
)=\pa^\a\phi\pa^\b\phi-\frac{1}{2}g^{\a\b}(\pa^\l\phi\pa_\l\phi-2V(\phi)),
\ee and we have the following result.
\begin{theorem}\label{2}
The energy density $P^\a n_\a$ of the scalar field $\phi$ with
potential $V(\phi)$ is positive when   $V(\phi)>0$.
\end{theorem}
\begin{proof}
The proof, which we  give  here for a general
scalar field potential $V(\f )$, is a direct adaptation with  slight modifications of that
found in \cite{ch-mo00}, p. 88, for a power-law potential. Using
Eq. (\ref{t}) we calculate
\be
P^\a n_\a =-\frac{1}{2}X^\a n_\a\pa^\lambda\f\pa_\lambda\f
+X^\a\pa_\a\f n^\b\pa_\b\f +X^\a n_\a V(\f ) .
\ee
We define the quadratic form
\be
\gamma^{\l\m}=-g^{\l\m}X^\a n_\a +(X^\l n^\m +X^\m n^\l )
\ee
and then we find that
\be
P^\a n_\a =-\frac{1}{2}X^\a n_\a g^{\l\m}\pa_\l\f\pa_\m\f +\frac{1}{2}(X^\l n^\m +X^\m n^\l )
\pa_\l\f\pa_\m\f +X^\a n_\a V(\f ).
\ee
This means that
\be\label{pn}
P^\a n_\a =\frac{1}{2}\gamma^{\l\m}\pa_\l\f\pa_\m\f +X^\a n_\a V(\f ).
\ee
Since $\mathcal{M}_t$ is a $t=\mathrm{const.}$ hypersurface,
we can choose coordinates such that $X^0=1, X^i=0, n_i=0$. Then
$n_0=(g^{00})^{-1/2}$, $n^i=g^{i0}(g^{00})^{-1/2}$ and so
it follows that the quadratic form $\gamma$ is positive
definite,
\be
\gamma^{00}=g^{00}n_{0},\quad\gamma^{i0}=0\quad\gamma^{ij}=-g^{ij}n_{0},
\ee
(recall signature of $g_{ij}$ is $(-\cdots -)$). We then find that
\bq
P^\a n_\a &=&\frac{1}{2}\left(\gamma^{00}\pa_0\f\pa_0\f
+2\gamma^{i0}\pa_i\f\pa_0\f +\gamma^{ij}\pa_i\f\pa_j\f\right) +n_0
V(\f )\nonumber\\
&=&\frac{1}{2}\left( n_0
g^{00}\dot{\f}^{2}-g^{ij}n_{0}\pa_i\f\pa_j\f\right) +n_0 V(\f
)\nonumber
\eq
and therefore we conclude that the energy density $ P^\a n_\a $ is
positive whenever  $V(\phi)>0$. This concludes the proof.
\end{proof}

To end this Section, for the following simple application of the
preceding developments we restrict attention to $n=4$ spacetime
dimensions although everything we do becomes valid with minor
modifications to arbitrary $n$. The following notation for
conformally related quantities is used: Let $g$ and $\tilde g$ be
two conformal metrics, $\tilde g =\Omega^{2} g$, on the manifold
$\mathcal{V}$. This means that in two \emph{orthonormal} moving
frames, $\theta^\a$ and $\tilde\theta^\a$, the two conformal metrics
satisfy \be \tilde g
=\eta_{\a\b}\tilde\theta^\a\tilde\theta^\b,\quad
g=\eta_{\a\b}\theta^\a\theta^\b\quad\textrm{and}\quad\tilde\theta^\a=\Omega^{-1}\,\theta^\a
, \ee with $\eta_{\a\b}=\textrm{diag} (+,-\cdots -)$ being the flat
metric. Setting $\Omega^{2} =e^{\f}$ we see that
$\tilde\theta^\a=e^{-\f/2}\theta^\a$ and obviously
$\tilde\theta_\a=e^{\f/2}\theta_\a$. The same rules are true for any
1-form or vectorfield  on $\mathcal{V}$. Consider now the
$f(R)$-\emph{vacuum} equations, \be\label{f of r vac} L_{\a\b}\equiv
f'R_{\a\b}-\frac{1}{2}g_{\a\b}f-\na_\a\na_\b
f'+g_{\a\b}\,\square\,_{g}f'=0, \ee where the left hand side
satisfies the conservation identities (cf. \cite{ed23}, p. 140)
\be\label{cons id} \na_\a L^{\a\b}=0. \ee Then we conformally
transform from  $(\mathcal{V},g)$ to the Einstein frame
$(\mathcal{V},\tilde{g})$, according to the prescription given in
\cite{con}, that is, we set \be\label{ct} \f =\ln f', \ee to obtain
the Einstein equations with a scalar field `matter source' of
potential $V(\phi)=(1/2)(f')^{-2}(Rf'-f)$ and energy-momentum tensor
given by Eq. (\ref{t}): \be\label{ceq}
\tilde{G}_{\a\b}=\tilde{T}_{\a\b}(\f ). \ee In this case we conclude
that the field $\f$ is conserved, i.e., \be \tilde{\na}_\a
\tilde{T}^{\a\b}(\f )=0, \ee and, since \be\label{sw0}
\tilde{\na}_\a \tilde{T}^{\a\b}(\f )=\pa^\b\f
(\tilde{\na}_\a\pa^\a\f+V'), \ee we find that the $\f$-field is a
scalar field satisfying the wave equation \be\label{sw1}
\tilde{\na}_\a\pa^\a\f+V'=0. \ee Further from Theorem \ref{1} we
have the following result.
\begin{corollary}
The slice energy of the scalar field $\f$ generated by the conformal transformation (\ref{ct})
to the Einstein frame of the $f(R)$-\emph{vacuum} equations (\ref{f of r
vac}) relative to a Killing vectorfield of $\tilde g$ is
conserved, i.e.,
\be
\tilde{E}_{t}(\f )=\int_{\mathcal{M}_{t}}
\tilde{P}^{\a}\tilde{n}_{\a}d\tilde{\m}_{t}=\mathrm{const,}
\ee
with $d\tilde{\m}_{t}$ being the volume element of $\tilde{g}_{t}$.
\end{corollary}

Secondly  from Theorem \ref{2} we have:
\begin{corollary}
For all $f(R)$-vacuum theories (\ref{f of r vac}) with a positive potential
in the Einstein frame the energy density $\tilde{P}^\a \tilde{n}_\a$
 of $\f$  is positive.
\end{corollary}
Examples of theories in the last Corollary include, for instance,
the choice $f(R)=R+\a R^2, \a >0$.

\section{$f(R)$-matter systems}
Suppose now that we start by coupling a matter field $\psi$ to the
geometry in  $(\mathcal{V},g)$ via the $f(R)$-matter field
equations
\be
f'R_{\a\b}-\frac{1}{2}g_{\a\b}f-\na_\a\na_\b
f'+g_{\a\b}\,\square\,_{g}f'=T_{\a\b}(\psi).
\ee
Because of the conservation identities (\ref{cons id}), the field
$\psi$ satisfies the conservation laws
\be\label{cons laws}
\na_\a T^{\a\b}(\psi)=0.
\ee
Then, if we conformally transform from  $(\mathcal{V},g)$ to the Einstein frame
$(\mathcal{V},\tilde{g})$ according to (\ref{ct}),  in place of
equations (\ref{ceq}) we obtain
\be\label{ceq1}
\tilde{G}_{\a\b}=\tilde{T}_{\a\b}(\f
)+\tilde{T}_{\a\b}(\tilde{\psi} ),
\ee
where now the whole tensor in the right-hand-side is conserved,
namely
\be\label{3.7}
\tilde{\na}_\a\left(\tilde{ T}^{\a\b}(\f )+\tilde{T}^{\a\b}(\tilde{\psi} )\right)=0,
\ee
but the two components are \emph{not} conserved separately, that is
\be
\tilde{\na}_\a \tilde{T}^{\a\b}(\f )\neq 0
\ee
and
\be\label{3.9}
\tilde{\na}_\a \tilde{T}^{\a\b}(\tilde{\psi} )\neq 0,
\ee
unless the conservation equations (\ref{cons laws}) for the field $\psi$ are
 conformally invariant (conditions for this are given in \cite{wa84}, p. 448).
 This result (already given in Ref. \cite{co93}) indicates that in higher order gravity theories
there must be a generic, nontrivial $\phi -\tilde\psi$ interaction
between the  matter field $\tilde{\psi}$ and the  $\phi$-field, and an associated exchange of energy between
$\phi$ and $\tilde\psi$.

Writing Eq. (\ref{trans-eq}) for the scalar field $\f$ and substituting
for the last term
in the right-hand-side from Eq. (\ref{3.7}) we find the \emph{general
energy transport equation} in the Einstein frame,
\be\label{en1dust}
E_{t_{1}} (\f )-E_{t_{0}} (\f )=\frac{1}{2}\int_{t_{0}}^{t_{1}}\int_{\mathcal{M}_{t}}
\tilde{T}^{\a\b}(\f)(\tilde{\na}_{\a}\tilde{X}_{\b}+\tilde{\na}_{\b}\tilde{X}_{\a})d\tilde{\mu}
-\int_{t_{0}}^{t_{1}}\int_{\mathcal{M}_{t}}
\tilde{X}_{\b}\tilde{\na}_{\a}\tilde{T}^{\a\b}(\tilde{\psi}
)d\tilde{\mu},
\ee
with $ d\tilde{\m}$ being the volume element of $\tilde{g}$.

This result is only symbolic and has to be augmented by precise
equations satisfied by the fields.
Since the
stress tensor of the $\phi$-field is not separately conserved, it follows that
the $\phi$-field will not satisfy the usual scalar wave equation,
$\widetilde{\square}_{\tilde{g}}\f +V'(\f )=0$, but this equation will  in
general contain new terms. Similarly for the `ordinary matter'  $\tilde\psi$-field,
whatever its form (scalar field,  Maxwell,  a fluid etc), its
field equations have new terms indicating the $\phi -\tilde\psi$
interaction and associated energy exchange. For instance, if the $\tilde\psi$ field is another scalar
field, then the equations satisfied by its conformal transform,
$\tilde\psi$,  have a general form of the type
$
\widetilde\square\tilde\psi +h(\f)\,\pa_\a\f\,\pa^\a\tilde\psi =0,
$
where $h(\f)$ is a smooth function of $\f$ (often exponential).

To study this interaction and the associated energy exchange between
$\phi$ and $\tilde{\psi}$ more closely we give some concrete
examples. Suppose firstly that $\psi$ is  a dust cloud on
$(\mathcal{V},g)$  with 4-velocity $V_\a$ and stress tensor \be
T_{\a\b,\,\textrm{dust}}=\rho V_\a V_\b, \ee satisfying the
$f(R)$\emph{-dust}  equations in the Jordan frame, namely
\be\label{f-dust} f'R_{\a\b}-\frac{1}{2}g_{\a\b}f-\na_\a\na_\b
f'+g_{\a\b}\,\square\,_{g}f'=\rho V_\a V_\b.  \ee Then
\be\label{cons dust} \na_\a (\rho V^\a V^\b)=0, \ee and it is
obvious that here the dust streamlines are geodesics, that is $V^\a$
is the tangent vectorfield to the geodesics. After the conformal
transformation  we find \be\label{ein-frame-dust}
\tilde{G}_{\a\b}=\tilde{T}_{\a\b}(\f)+\tilde{\rho}\, \tilde{V}_\a
\tilde{V}_\b, \ee with  $\tilde{T}_{\a\b}(\f)$ given by (\ref{t})
with tildes where appropriate, and with\be \tilde{V_{a}}=e^{-\f/2}
V_{a},\quad\tilde{\rho}=e^{-2\f}\rho. \ee (We have set $\tilde\rho
=\Omega^{-4}\rho$ and since $\Omega^2=e^\f$, $\Omega^{-4}
=e^{-2\f}$.)

What is the field equation satisfied by the scalar
field $\f$? From Eq. (\ref{ein-frame-dust}) the divergence of the
stress tensor of $\f$ is minus that of the dust, but \be
\tilde{\na}_\a (\tilde{\rho}\, \tilde{V}^\a \tilde{V}^\b )= \na_\a
(\tilde{\rho}\, \tilde{V}^\a \tilde{V}^\b )+A^\a_{\a\g}
\tilde{\rho}\, \tilde{V}^\g \tilde{V}^\b+A^\b_{\a\g} \tilde{\rho}\,
\tilde{V}^\a \tilde{V}^\g, \ee
 where
 \be
A^\a_{\b\g}=\frac{1}{2}\left(\delta^{\a}_{\b}\pa_{\g}\f
+\delta^{\a}_{\g}\pa_{\b}\f -g_{\b\g}g^{\a\d}\pa_{\d}\f\right).
 \ee
From these equations and Eq. (\ref{cons dust}) we deduce the
modified scalar field equation in the form
\be\label{sc field eqn}
\pa^{\b}\f(\widetilde{\square}\f+V')+\frac{1}{2}
\tilde{\rho}\, \tilde{V}^\a \tilde{V}^\b
\pa_\a\f -\frac{1}{2}\tilde\rho\pa^\b\f=0.
\ee
Another way to derive  the scalar field equation is as follows.
Since
\be\label{3.16}
\tilde{\na}_\a \tilde{T}^{\a\b}_{\textrm{dust}}=\tilde{\na}_\a
(\tilde{\rho}\, \tilde{V}^\a \tilde{V}^\b )=
\tilde{V}^\b\tilde{\na}_\a (\tilde{\rho}\, \tilde{V}^\a
)+\tilde{\rho}(\tilde{\na}_\a \tilde{V}^\b )\tilde{V}^\a ,
\ee
and,  since $\tilde{V}_\b \tilde{V}^\b =1$, if we multiply Eq. (\ref{sw0}) by $\tilde{V}_\b$  and use the fact that
the divergence of the right hand side of
Eq. (\ref{ein-frame-dust}) is zero to arrive at the following equation for the scalar field $\f$
in the Einstein frame, namely,
\be\label{sc field eqn1}
\pa^{\b}\f(\widetilde{\square}\f+V')+
\tilde{V}^{\b}\,\tilde{\na}_{\a}(\tilde{\rho}\,\tilde{V}^{\a})
+\tilde{\rho}\tilde{V}^\a \tilde{\na}_\a \tilde{V}^\b =0.
\ee
Recalling that dust matter follows geodesics on the original Jordan
frame, $V^\a\na_\a V^\b=0$, and taking it \emph{as a working hypothesis} that the same is true in the
conformally related Einstein frame, we find that the last two terms
in this equation are equal to the last two
terms in Eq. (\ref{sc field eqn}) and so we conclude that Eq. (\ref{sc field
eqn1}) provides an equivalent form of Eq. (\ref{sc field eqn}).

We note that only in the very special case where we impose the  constraint
\be\label{3.18}
\tilde{V}_{\b}=\pa_{\b}\f,
\ee
which implies some sort of `alignment' between the dust component and the
scalar field,  does the scalar field equation (\ref{sc field eqn}) becomes the standard one, namely,
\be
\widetilde{\square}\f+V' =0.
\ee

We now study the behaviour of the total slice energy of the system comprised of $\f$ and
the dust component.  We choose $V=X$ so that
\be\label{3.11}
P^\a n_\a =X_\b n_\a\rho V^\a V^\b=\rho V^\a n_\a.
\ee
Hence,  applying Stokes' theorem we obtain
\be\label{stokes}
\int_{\mathcal{K}}\tilde{\na}_\a (\tilde{\rho} \tilde{V}^\a )d\tilde\mu
=\int_{\pa\mathcal{K}}\tilde{\rho} \tilde{V}^\a \tilde{n}_\a d\tilde\sigma .
\ee
Therefore Eq. (\ref{en1dust}) becomes
\bq
E_{t_{1}} (\f )-E_{t_{0}} (\f )
&=&\frac{1}{2}\int_{t_{0}}^{t_{1}}\int_{\mathcal{M}_{t}}
\tilde{T}^{\a\b}(\f )(\tilde{\na}_{\a}\tilde{V}_{\b}+\tilde{\na}_{\b}\tilde{V}_{\a})d\tilde{\mu}
\nonumber\\&-&
\left[\int_{\mathcal{M}_{t_{1}}}\tilde\rho\tilde{V}^\a \tilde{n}_\a d\tilde\m_{t_{1}}
-
\int_{\mathcal{M}_{t_{0}}}\tilde\rho\tilde{V}^\a \tilde{n}_\a d\tilde\m_{t_{0}}\right]
\nonumber\\
&=&\frac{1}{2}\int_{t_{0}}^{t_{1}}\int_{\mathcal{M}_{t}}
\tilde{T}^{\a\b}(\f )(\tilde{\na}_{\a}\tilde{V}_{\b}+\tilde{\na}_{\b}\tilde{V}_{\a})d\tilde{\mu}
+\int_{\mathcal{M}_{t_{0}}}\tilde\rho\tilde{V}^\a \tilde{n}_\a
d\tilde\m_{t_{0}}\nonumber\\&-&\int_{\mathcal{M}_{t_{1}}}\tilde\rho\tilde{V}^\a \tilde{n}_\a
 d\tilde\m_{t_{1}},
 %\nonumber\\
\eq
or
\be\label{3.26A}
E_t (\f ) +E_t (\mathrm{dust}) =E_0 (\f ) +E_0 (\mathrm{dust})
+\frac{1}{2}\int_{t_{0}}^{t_{1}}\int_{\mathcal{M}_{t}}
\tilde{T}^{\a\b}(\f )(\tilde{\na}_{\a}\tilde{V}_{\b}+
\tilde{\na}_{\b}\tilde{V}_{\a})d\tilde{\mu},
\ee
where by definition and Eq. (\ref{3.11}), for any $t$,
\be
E_t (\mathrm{dust})=\int_{\mathcal{M}_{t}}\rho V^\a n_\a d\m_{t}.
\ee
We see that the last term in Eq. (\ref{3.26A}) can be zero only
when $V$ is a Killing vectorfield.
%Since $\tilde{T}_{\a\b}=\tilde{\rho} \tilde{V}_\a \tilde{V}_\b,$ we have
%\be
%\tilde{T}^{\a\b}(\tilde{\na}_{\a}\tilde{V}_{\b}+\tilde{\na}_{\b}\tilde{V}_{\a})=
%\tilde\rho\tilde V^{\a}\tilde V^{\b}\tilde\na_{\a}\tilde V_{\b}+\tilde\rho\tilde
%V^{\b}\tilde
%V^{\a}\tilde\na_{\b}\tilde V_{\a}=0,
%\ee
%always, irrespective of whether or not $V$ is a Killing
%vectorfield. The same result evidently holds also in the Jordan frame.
We therefore arrive at the following result about the total slice energy with respect to the
fluid itself.
\begin{theorem}\label{thm1}
The total  slice energy with respect to the timelike
vectorfield $\tilde V$, tangent to the dust timelines, of the scalar
field-dust system satisfying the field equations (\ref{f-dust}),
satisfies
\be\label{3.26AA}
E_t (\f +\mathrm{dust}) =E_0 (\f + \mathrm{dust})
+\frac{1}{2}\int_{t_{0}}^{t_{1}}\int_{\mathcal{M}_{t}}
\tilde{T}^{\a\b}(\f )(\tilde{\na}_{\a}\tilde{V}_{\b}+
\tilde{\na}_{\b}\tilde{V}_{\a})d\tilde{\mu}.
\ee
In particular the slice energy of the scalar
field-dust system is conserved when $\tilde V$ is a Killing vectorfield of $\tilde g$.
\end{theorem}
We also conclude that the property of the conservation of slice
energy for dust is a conformal invariant. However, when $V$ is not a Killing vectorfield, we
see that there is a nontrivial contribution to the slice energy
coming from the combined effect of the stress tensor of the scalar field generated by the
conformal transformation coupled to the non-stationarity of the spacetime
due to the lack of a Killing vector.
Note that this contribution is also
nonzero even in the special case that Eq. (\ref{3.18}) is assumed for
in
that case the first term in $\tilde{T}^{\alpha \beta }(\phi )
(\tilde{\nabla}_{\alpha }
\tilde{V}_{\beta }+\tilde{\nabla}_{\beta }\tilde{V}_{\alpha})$ is
zero because $\tilde{V}^{\alpha }\tilde{V}^{\beta }(\tilde{\nabla}_{\alpha }%
\tilde{V}_{\beta }+\tilde{\nabla}_{\beta}\tilde{V}_{\alpha})=0$,
but the whole combination is still not zero as there are additional
terms coming from the contributions of the other terms in Eq.
(\ref{t}) (unless the fluid satisfies an extra condition -- see below).

We now proceed to see how this result changes when we assume
that $(\mathcal{V},g)$ is filled with a perfect fluid. With our
conventions the stress tensor of  a perfect fluid with energy density $\rho$ and
pressure density $p$ is $T_{\a\b}=(\rho +p)V_\a V_\b -pg_{\a\b}$
and the fluid satisfies the field equations
\be\label{f(R)-fluid}
f'R_{\a\b}-\frac{1}{2}g_{\a\b}f-\na_\a\na_\b
f'+g_{\a\b}\Box_{g}f'=(\rho +p)V_\a V_\b -pg_{\a\b}.
\ee
In this case, because the energy-momentum vector $P^\a =\rho V^\a$, we have $P^\a n_\a =
\rho V^\a n_\a$ and so the slice energy with
respect to the timelike vectorfield $\tilde{V}$ in the Einstein frame is again
%($\tilde{p}=e^{-2\f}p$)
\be
E_{t}(\mathrm{fluid})=\int_{\mathcal{M}_{t}}\tilde{\rho}\tilde{V}^\a \tilde{n}_\a d\tilde\m_{t}.
\ee
Then
\be\label{1a}
\tilde\na_\a \tilde T^{\a\b}_{\textrm{fluid}}=\tilde\na_\a
\left[ (\tilde\rho +\tilde p)\tilde V^\a \right] \tilde V^\b +(\tilde\rho
+\tilde p)\tilde V^\a \tilde\na_\a  \tilde V^\b -\tilde\na^\b\tilde
p.
\ee
Since $\tilde V_\a\tilde V^\a =1$, we have $\tilde V_\a\tilde\na_\b\tilde V^\a
=0$ and so on multiplication of Eq. (\ref{1a}) by $\tilde V_\b$ we find
\bq\label{3.25}
\tilde V_\b\tilde\na_\a\tilde T^{\a\b}_{\mathrm{fluid}}
&=&\tilde\na_\a
\left[ (\tilde\rho +\tilde p)\tilde V^\a \right] -
\tilde V^\b \pa_\b\tilde p\nonumber\\
&=&\tilde p\,\tilde\na_\a\tilde V^\a
+\tilde\na_{\a}(\tilde\rho\,\tilde{V}^{\a}).
\eq
Integrating Eq. (\ref{3.25}) on the spacetime slab $\mathcal{D}$ and
using Eq. (\ref{stokes}) to re-express the $\rho$-term in (\ref{3.25}) we have
\be
\int_{t_{0}}^{t_{1}}\int_{\mathcal{M}_{t}}
-\tilde V_{\b}\tilde\na_{\a}\tilde T^{\a\b}_{\mathrm{fluid}}d\tilde\mu
=\int_{t_{0}}^{t_{1}}\int_{\mathcal{M}_{t}}-\tilde p\tilde\na^\a\tilde V_\a d\tilde\m
-\left[\int_{\mathcal{M}_{t_{1}}}\tilde\rho d\tilde\m_{t_{1}}-
\int_{\mathcal{M}_{t_{0}}}\tilde\rho d\tilde\m_{t_{0}}\right]
\ee
and therefore we obtain from (\ref{en1dust}) the general
energy transport equation in the form
\bq
E_{t_{1}} (\f )-E_{t_{0}} (\f ) &=&\frac{1}{2}\int_{t_{0}}^{t_{1}}\int_{\mathcal{M}_{t}}
\tilde{T}^{\a\b}(\f)(\tilde{\na}_{\a}\tilde{V}_{\b}+\tilde{\na}_{\b}\tilde{V}_{\a})d\tilde{\mu}
\nonumber\\
&-&\int_{t_{0}}^{t_{1}}\int_{\mathcal{M}_{t}}\tilde
p\,\tilde\na^\a\tilde V_\a
d\tilde\m +E_{t_{0}} (\mathrm{fluid})-E_{t_{1}} (\mathrm{fluid}).
\eq
We thus arrive at the following result.
\begin{theorem}
The total slice energy  of the scalar
field -- perfect fluid system satisfying the field equations (\ref{f(R)-fluid}),
depends upon the integrated pressure according to the formula
\be\label{en1fluid}
E_{t_{1}} (\f +\mathrm{fluid})=E_{t_{0}} (\f +\mathrm{fluid})
+\frac{1}{2}\int_{t_{0}}^{t_{1}}\int_{\mathcal{M}_{t}}
\tilde{T}^{\a\b}(\f)(\tilde{\na}_{\a}\tilde{V}_{\b}+\tilde{\na}_{\b}\tilde{V}_{\a})d\tilde{\mu}
-\int_{t_{0}}^{t_{1}}
\int_{\mathcal{M}_{t}}\tilde p\,\tilde\na^\a \tilde V_\a d\tilde\m.
\ee
In particular, the slice energy
is conserved when  $\tilde V$ is a Killing vectorfield for $\tilde g$ (stationary spacetime).
\end{theorem}
When $V$ is not a Killing vectorfield, this slice energy is not generally
conserved and this is true even in the special case of a
fluid with zero expansion, $\tilde\na^\a \tilde V_\a =0$, for which the
last term in Eq. (\ref{en1fluid}) is zero. In this case the term
depending on the scalar field continues to have a nonzero
contribution to the total slice energy. This term is given by
\be
T^{\a\b}(\f)(\tilde{\na}_{\a}\tilde{V}_{\b}+\tilde{\na}_{\b}\tilde{V}_{\a})=
\pa^\a\f\pa^\b\f (\tilde{\na}_{\a}\tilde{V}_{\b}+\tilde{\na}_{\b}\tilde{V}_{\a})
-\tilde\na_\a \tilde V^\a (\pa^\l\f\pa_\l\f -2V(\f))
\ee
and so  Eq. (\ref{en1fluid}) becomes
\bq\label{en1fluid1}
E_{t_{1}} (\f +\mathrm{fluid})&=&E_{t_{0}} (\f +\mathrm{fluid})
+\frac{1}{2}\int_{t_{0}}^{t_{1}}\int_{\mathcal{M}_{t}}
\pa^\a\f\pa^\b\f(\tilde{\na}_{\a}\tilde{V}_{\b}+\tilde{\na}_{\b}\tilde{V}_{\a})d\tilde{\mu}
\nonumber\\
&-&\int_{t_{0}}^{t_{1}}
\int_{\mathcal{M}_{t}}(\frac{1}{2}\pa^\l\f\pa_\l\f
+\tilde p -V(\f))\,\tilde\na^\a \tilde V_\a d\tilde\m.
\eq
We conclude that the only other possible case for which we have slice
energy conservation is when $V$ is not a Killing  vectorfield for $\tilde
g$,
but the alignment condition (\ref{3.18})
holds (making the middle term in Eq. (\ref{en1fluid1}) equal to zero)
and the fluid has in addition zero expansion (last term in Eq. (\ref{en1fluid1}) is zero).

\section{Discussion}
The results of this paper allow us to make some comments concerning the problem of deciding
which of the two frames (or metrics), Jordan or Einstein, is the
physical one, meaning in which of the two representations of the
dynamics test particles follow geodesics (assuming the validity of
the principle of equivalence). In the case where test particles
follow geodesics in both frames, one says that the two conformally
related frames are \emph{physically equivalent}. The main results of
 Section 3, in particular Eq. \ref{3.26AA} (as well as its
pressure extension - Eq. \ref{en1fluid}), were proved under the
implicit assumption that the vectorfield $V^a$ of the dust
streamlines generates  geodesics in both the Jordan frame \emph{and}
the conformally related Einstein frame.

But does $\tilde{V}%
^{\alpha }$ always generate a geodesic in the latter frame so that $\tilde{V}_{\beta }%
\tilde{\nabla}_{\alpha }\tilde{T}_{dust}^{\alpha \beta }=\tilde{\nabla}%
_{\alpha }(\tilde{\rho}\tilde{V}^{\alpha })$ (cf. Eq. \ref{3.16})?
In general, it will not do so and the two frames will not be
physically equivalent. In this case, we have \be
\tilde{V}_\b\tilde{\na}_\a
\tilde{T}^{\a\b}_{\textrm{dust}}=\tilde{\na}_\a (\tilde{\rho}\,
\tilde{V}^\a )\tilde{V}_\b \tilde{V}^\b +\tilde{\rho}
(\tilde{\na}_\a \tilde{V}^\b )\tilde{V}^\a \tilde{V}_\b, \ee so that
the result of Theorem \ref{thm1} becomes,
\begin{eqnarray}\label{3.26AAA} E_t (\f +\mathrm{dust}) &=&E_0 (\f +
\mathrm{dust})
+\frac{1}{2}\int_{t_{0}}^{t_{1}}\int_{\mathcal{M}_{t}}
\tilde{T}^{\a\b}(\f )(\tilde{\na}_{\a}\tilde{V}_{\b}+
\tilde{\na}_{\b}\tilde{V}_{\a})d\tilde{\mu}\nonumber\\&-&\int_{t_{0}}^{t_{1}}\int_{\mathcal{M}_{t}}\tilde{\rho}
(\tilde{\na}_\a \tilde{V}^\b )\tilde{V}^\a \tilde{V}_\b
d\tilde{\mu},\nonumber\\
&=&E_0 (\f + \mathrm{dust})
+\frac{1}{2}\int_{t_{0}}^{t_{1}}\int_{\mathcal{M}_{t}}
\left(\tilde{T}^{\a\b}(\f )-\tilde{\rho}\tilde{V}^\a
\tilde{V}^\b\right)(\tilde{\na}_{\a}\tilde{V}_{\b}+
\tilde{\na}_{\b}\tilde{V}_{\a})d\tilde{\mu}.\nonumber\\
\end{eqnarray}
We may therefore conclude that the expressions for the total slice
energy in the two situations considered here, namely, when test
particles follow geodesics in both metrics $g,\tilde{g}$ (cf.  Eq.
\ref{3.26AA}), or only in the original Jordan frame metric $g$ (cf.
Eq. \ref{3.26AAA}), are different and in the latter case there is an
extra term contributing to the total energy (i.e., the dust term in
the integrand in the last term in Eq. \ref{3.26AAA}). This
additional contribution will appear as a measurable quantity which,
if measured to be nonzero, will lead us to conclude that the two
conformally related frames cannot be physically indistinguishable.

\section*{Acknowledgements}
We are indebted to Y. Choquet-Bruhat for her encouragement and for
offering her precious time and opinion so
willingly. These have led to a substantial improvement of this work.
We also thank J. Miritzis for reading this paper and
P.G.L. Leach for comments on a preliminary version of this manuscript.

\end{document}